\documentclass[superscriptaddress,showpacs,floatfix,longbibliography]{revtex4-1}
\usepackage{amsmath,amssymb,graphicx}
\usepackage{wasysym}
\usepackage[utf8]{inputenc}
\usepackage[T1]{fontenc}
\usepackage{xcolor}

\usepackage{rotating} 

\IfFileExists{newtxtext.sty}
   {\usepackage{newtxtext,newtxmath}}
   {\IfFileExists{stix.sty}
      {\usepackage{stix}}
      {\IfFileExists{mathptmx.sty}
      {\usepackage{mathptmx}}{} } }

\usepackage{textcomp}

\usepackage{bm}

\IfFileExists{siunitx.sty}{\usepackage{booktabs,siunitx}}{}

\pdfoutput=1
\usepackage{color}
\definecolor{LinkColor}{rgb}{0.256,0.439,0.588}
\usepackage{hyperref}

\newcommand{\Z}{\mathbb{Z}}

\usepackage{pifont}

\newcommand{\La}{\line (1,0  ){12}}
\newcommand{\Lb}{\line (3,5 ){6}}

\newcommand{\Ld}{\line (-1,0){12}}
\newcommand{\Le}{\line (-3,-5){6}}
\newcommand{\Lf} {\line(3,-5){6}}
\newcommand{\C} {\circle*{4}}

\newcommand{\LaT}{\rule[-1pt]{0.4cm}{0.2em}}  
\newcommand{\LdT}{\rule[-1pt]{0.4cm}{0.2em}}  
\newcommand{\LbT}{\rotatebox{60}{\rule[-1pt]{0.4cm}{0.2em}}}  
\newcommand{\LeT}{\rotatebox{60}{\rule[-1pt]{0.4cm}{0.2em}}}  

\newcommand{\pA}{\put(-6,-10)}
\newcommand{\pB}{\put(6,-10)}
\newcommand{\pC}{\put(12,0)}

\newcommand{\pZ}{\put(0,0)}

\newcommand{\pAT}{\put(-6.8,-10)} 
\newcommand{\pBT}{\put(5.2,-10)}  

\newcommand{\rhomb}{
  \pA{\C}\pB{\C}\pZ{\C}\pC{\C}
 }

\newcommand{\rhombH}{
  \begin{picture}(22,10)(-8,-6)
    \pA{\LaT}\pB{\Lb}\pZ{\Le}\pZ{\LdT}\pZ{\Lf}
    \rhomb
  \end{picture}
}
\newcommand{\rhombV}{
  \begin{picture}(22,10)(-8,-6)
    \pA{\La}\pBT{\LbT}\pAT{\LeT}\pC{\Ld}\pZ{\Lf}
    \rhomb
  \end{picture}
}

\usepackage{lineno}
\begin{document}

\title{Topological phase transition and single/multi anyon dynamics of $Z_2$ spin liquid}

\author{Zheng Yan}
\affiliation{Department of Physics and HKU-UCAS Joint Institute of Theoretical and Computational Physics, The University of Hong Kong, Pokfulam Road, Hong Kong}
\author{Yan-Cheng Wang}
\affiliation{School of Materials Science and Physics, China University of Mining and Technology, Xuzhou 221116, China}
\author{Nvsen Ma}
\affiliation{School of Physics, Key Laboratory of Micro-Nano Measurement-Manipulation and Physics, Beihang University, Beijing 100191, China}
\author{Yang Qi}
\email{qiyang@fudan.edu.cn}
\affiliation{State Key Laboratory of Surface Physics, Fudan University, Shanghai 200433, China}
\affiliation{Center for Field Theory
	and Particle Physics, Department of Physics, Fudan University,
	Shanghai 200433, China}
\affiliation{Collaborative Innovation Center of Advanced
	Microstructures, Nanjing 210093, China}

\author{Zi Yang Meng}
\email{zymeng@hku.hk}
\affiliation{Department of Physics and HKU-UCAS Joint Institute of Theoretical and Computational Physics, The University of Hong Kong, Pokfulam Road, Hong Kong}

\begin{abstract}
Among the quantum many-body models that host anyon excitation and topological orders, quantum dimer models (QDM) provide a unique playground for studying the relation between single-anyon and multi-anyon continuum spectra. However, as the prototypical correlated system with local constraints, the generic solution of QDM at different lattice geometry and parameter regimes is still missing due to the lack of controlled methodologies. Here we obtain, via the newly developed sweeping cluster quantum Monte Carlo algorithm, the excitation spectra in different phases of the triangular lattice QDM. Our results reveal the single vison excitations inside the $Z_2$ quantum spin liquid (QSL) and its condensation towards the $\sqrt{12}\times\sqrt{12}$ valence bond solid (VBS), and demonstrate the translational symmetry fractionalization and emergent O(4) symmetry at the QSL-VBS transition. We find the single vison excitations, whose convolution qualitatively reproduces the dimer spectra, are not free but subject to interaction effects throughout the transition. The nature of the VBS with its O(4) order parameters are unearthed in full scope. Our approach opens the avenue for generic solution of the static and dynamic properties of QDMs and has relevance towards the realization and detection of fractional excitations in programmable quantum simulators.
\end{abstract}


\maketitle

\section*{Introduction}
Fractionalized anyon excitations are among the most important features of topologically ordered phases, a class of phases beyond the Landau paradiam of classifying phases with symmetry breaking~\cite{Wen2019}.
The fractionalized nature of these anyon excitations renders that they cannot be created or annihilated individually by physical probes.
This phenomenon is both a blessing and a curse: it reflects the topological nature of the excitations and the phase, but also obscures any direct detection of single anyon excitations.
Instead, they can only be observed indirectly from a multi-particle continuum of spectral functions.
For example, a continuum in inelastic neutron scattering spectrum is often used as a signature to detect quantum spin liquids with fractionalized spin excitations, which is considered as a two-spinon continuum~\cite{HanTH12,WeiYuan2017,feng2018claringbullite,wen2019the,YuanWei2020}.
Consequently, understanding the relation between physical spectra and underlying single-anyon excitation is an essential question in the study of topologically ordered phases.

In the theoretical study of topologically ordered phases including the quantum spin liquid (QSL)~\cite{YiZhou2017,Broholm2020}, one usually relies on approximate tools to model the fractionalized excitations because they cannot be directly accessed in experiments and numerical simulations. In simple mean-field theories of QSL, as a physical probe excites a pair of fractionalized excitations, the corresponding spectrum is given by the convolution of spectra of the underlying anyons.
However, in realistic systems, this simple relation is modified by interactions between anyons, it is therefore important to know how much change has happened due to the interaction effect.

Quantum dimer models (QDM) ~\cite{Kivelson1987,Rokhsar1988} provide a unique playground for studying the relation between single-anyon and two-anyon spectra in QSLs.
Originally proposed to model the resonant valence bond state in high-$T_c$ superconductors~\cite{Baskaran1988} and frustrated magnets, it realizes a gapped $Z_2$ QSL at the exactly-solvable Rokhsar-Kivelson (RK) point~\cite{Rokhsar1988} if put on a nonbipartite lattice such as the triangle and the kagome~\cite{MoessnerSondhi2001a,MoessnerSondhi2001b,furukawa2007topological}.
Comparing to other models of QSLs, the QDMs are unique as the spinful excitations are absent in the Hilbert space due to the one-dimer-per-site constraint. This means that the spinon excitations in the $Z_2$ spin liquid are absent, leaving the visons as the only low-energy anyon excitations.
As a result, the spectrum of vison excitations can be directly measured in numerical simulations.
This feature of QDM allows one to compare the spectra of both the fractionalized single-vison excitations and the physical dimer-dimer correlations, which involves a pair of visons.
Although the ground state of QDM is exactly known at the RK point, the excited states are not exactly solvable due to interactions among the visons. 

Furthermore, away from the RK point, the QDM on the triangular lattice can be tuned into a $\sqrt{12}\times\sqrt{12}$ valence bond solid (VBS) phase~\cite{MoessnerSondhi2001a,MoessnerSondhi2001b}.
The phase transition is conjectured to be continuous and of the O(4) universality, driven by the condensation of visons~\cite{Ivanov2004,Ralko2005,Ralko2006,Ralko2007}.
Therefore, the QDM near this transition is an ideal system to study the spectral properties of anyon condensation, if there exist controlled theoretical and numerical methods.

\begin{figure*}[htp!]
	\centering
	\includegraphics[width=0.95\textwidth]{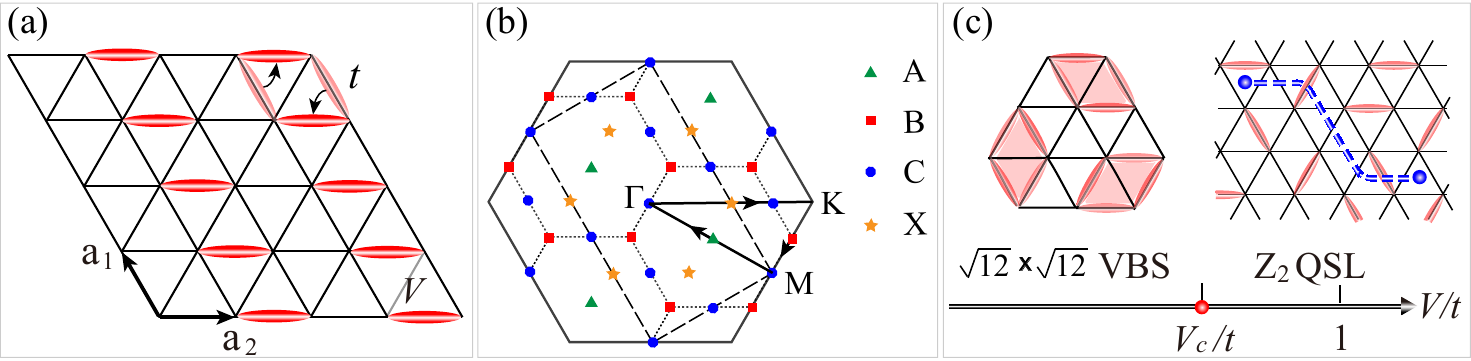}
	\caption{(a) The triangular lattice QDM. The two terms in the Hamiltonian Eq.~\eqref{eq:eq1} are depicted. The primitive vectors $\mathbf{a}_{1}$ and $\mathbf{a}_{2}$ are shown. The columnar reference dimer configuration for the measurement of vison correlations is also shown. (b) The solid hexagon and dashed rectangle are the Brilliouin zone (BZ) for the dimer and vison correlations, respectively, with $\Gamma - M - K - \Gamma$ the high symmetry path for the former and $A - B - C - X - A$ for the latter.  (c) Phase diagram of the triangular lattice QDM. The $V=1$ is the RK point and the $V_c = 0.85(5)$ separates the $Z_2$ QSL and $\sqrt{12}\times\sqrt{12}$ VBS phases~\cite{Ralko2007}. Approaching the $V_c$ from the $Z_2$ QSL, the dimer and vison-convolution spectral functions close gap at the $X$ ($M$) points and correspondingly the vison spectra close gap at $B$ point. (Insets) The enlarged unit cell of the  $\sqrt{12}\times\sqrt{12}$ VBS is shown, with its BZ the dashed hexagon in (b). A pair of visons in the QSL phase, with a string presenting an arbitrary path chosen to evalue the vison correlation function $C_{v}(r_{i,j})$.}
	\label{fig:fig1}
\end{figure*}

Recently, a new quantum Monte Carlo (QMC) scheme, the sweeping cluster method, is invented by the author~\cite{YanZheng2019a,YanZheng2019b,ZhenYan2020}. The method is able to keep track of the strict local constraint of dimer covering and at the same time perform Markov chain Monte Carlo (MC) in the space-time path integral such that both static and dynamic properties of the QDM can be obtained, only subject to finite system sizes. Therefore, it is different from the projection QMC employed in the previous literatures~\cite{Ivanov2004,Ralko2005,Ralko2006,Ralko2007}, where the interplay of quantum and thermal fluctuations of the QDM models is not present, and the computation complexity has been reduced such that larger system sizes can now be accessed (as will show later, the largest system size is three times larger than that in previous literature). The method has been applied to the square lattice QDM and a mixed phase separating columnar phase at strong dimer attraction and staggered phase at strong dimer repulsion are found~\cite{YanZheng2019b}. In this work, we further develop the method to study the static and dynamic properties of triangular lattice QDM.

The problem has a long and interesting history. From the work of Moessner-Sondhi~~\cite{MoessnerSondhi2001a}, one knows that from the mapping to frustrated Ising model on honeycomb lattice, the problem is in principle solvable via MC simulations on the frustrated Ising model, and a $\sqrt{12}\times\sqrt{12}$ VBS and a $Z_2$ QSL are suggested. Then in a series of works with zero temperature Green's function MC~\cite{Ivanov2004,Ralko2005,Ralko2006,Ralko2007}, the transition from the QSL to VBS, with the notion that the gap of topological vison excitations is closed at the transition is revealed, although the numerical method therein only work close to the RK point and zero temperature. Later, the dynamical dimer correlations at the RK point is presented in Ref.~\cite{Laeuchli2008}, taking the advantage that at the RK point, the quantum mechanics in imaginary time among the equally weighted dimer coverings is equivalent to a classical stochastic process~\cite{Henley2004}. Despite of these important progresses, the complete spectra of both dimer and vison excitations, not only the gap but also the spectral weight, and the complete understanding of the transition from $Z_2$ QSL to the  $\sqrt{12}\times\sqrt{12}$ VBS in terms of symmetry fractionalization of topological order, and the nature of the complex O(4) order parameter of the VBS, are not revealed. Here we try to answer these questions with unbiased QMC and symmetry analysis.

\section*{results}
{\noindent\it Model and Measurements.-} We study the QDM on triangular lattice with the Hamiltonian,
\begin{equation}
H=-t\sum_r \left(\left|\rhombV\right>\left<\rhombH\right| + h.c.\right)+V\sum_r\left(\left|\rhombV\right>\left<\rhombV\right|+\left|\rhombH\right>\left<\rhombH\right|\right)
\label{eq:eq1}
\end{equation}
where the sum runs over all plaquettes including the three possible
orientations. The kinetic term,  controlled by $t$, flips the
two dimers on every flippable plaquette, i.e., on every plaquette with two
parallel dimers, while the potential term $V$ describes interactions between nearest-neighbor dimers. Throughout the paper, we set $t=1$ as the energy unit and the inverse temperature $\beta=1/T$ with temperature scale also measured according to $t$.

Before the sweeping cluster QMC~\cite{YanZheng2019a,YanZheng2019b,ZhenYan2020}, one commonly employs the projector approaches to study QDMs, which includes the Green's function~\cite{Ivanov2004,Ralko2005,Ralko2006,Ralko2007} and diffusion MC schemes~\cite{OFS2005,OFS2006}. These projector methods obey the geometric constraints, but are not effective away from RK point~\cite{OFS2005walk} and only work at $T=0$. Also, there exists no cluster update for the projector methods to reduce the computational complexity. On the contrary, the sweeping cluster algorithm is based on path-integral in the world-line MC configurational space of all finite temperatures and features efficient cluster update for constrained systems. It is an general extension of the directed-loop algorithm~\cite{OFS2002,Alet2005a} for the D dimension classic dimer model~\cite{Alet2005b} to the quantum dimension of (D+1). Since our QMC works at finite temperature, we can also access the imaginary time correlation functions. And from here, we employ the stochastic analytic continuation (SAC) method~\cite{Sandvik1998a,Beach2004,Syljuasen2008,Sandvik2015,Qin2017,GYSun2018,Shao2017,Ma2018a,CKZhou2020,YCWangVestigial2020} to obtain the real frequency excitation spectra from their QMC imaginary time correspondance. The reliability of such QMC-SAC scheme has been extensively tested in quantum many-body systems, ranging from 1D Heisenberg chain~\cite{Sandvik2015} compared with Bethe ansatz, 2D Heisenberg model~\cite{Shao2017,CKZhou2020} compared with exact diagonalization, field theoretical analysis and neutron scattering of square lattice quantum magnet, $Z_2$ quantum spin liquid model with fractionalized spectra~\cite{GYSun2018,YCWangVestigial2020} compared with anyon condensation theory to quantum Ising model with direct comparison with neutron scattering and NMR experiments~\cite{Lih2020,ZHu2020}.

We compute three dynamical correlation functions. The first one is dimer correlation. The dimer operator $D_i=1$ or $0$ when there is a/no dimer on the link $i$. The dimer correlation function is defined as $C_d(r_{i,j},\tau) =\sum_{i,j} \langle D_i(\tau)D_j(0)\rangle-\langle D_i\rangle^2$, and $C_d(\mathbf{q},\tau)$ through the Fourier transformation, then the excitation spectrum $C_d(\mathbf{q},\omega)$ via SAC.

The second one is vison correlation. Visons ($V_i$) live in the centre of triangle plaquettes and they must arise in pairs, as shown in the right inset of Fig.~\ref{fig:fig1} (c). The correlation function is defined as $V_iV_j=(-1)^{N'_{P_{ij}}}$ where $N'_{P_{ij}}$ is the number of dimers along the path $P_{ij}$ we chose between plaquettes $i$ and $j$ as shown in Fig.~\ref{fig:fig1} (c). It is clear that the value of $V_iV_j$ is path dependent. In order to eliminate this dependence, one can choose a reference configuration, and follow the same path $P_{ij}$ again to obtain another $N''_{P_{ij}}$ and then the new observable $N_{P_{ij}}=N'_{P_{ij}}-N''_{P_{ij}}$ is path independent. Then we redefine $C_v(r_{i,j},\tau) = \langle V_i(\tau)V_j(0) \rangle = \langle V_i(\tau)V_i(0)V_i(0)V_j(0) \rangle=\langle (-1)^{N_{H_t}+N_{P_{ij}}} \rangle$ where $N_{H_t}$ means the number of the $t$-term in Eq.~\eqref{eq:eq1} between $V_i(\tau)$ and $V_i(0)$. We choose the reference configuration as the columnar VBS shown in Fig.~\ref{fig:fig1} (a), which doubles the unit cell and the corresponding BZ under this reference (gauge choice) is the dashed rectangle with high symmetry points $A$, $B$ and $C$ in Fig.~\ref{fig:fig1} (b).

The last one is another "dimer", i.e., the vison-convolution correlation function. We denote this "dimer" - the vison-convolution (VC) operator - as $D^{vc}_{i}=V_{i_1}V_{i_2}d_{i}$. The idea is that if two visons are closest to each other, sharing the same link, then the $D^{vc}_i$ on link $i$ can be represented as the product of these two vison operators, with $i_1$ and $i_2$ the triangle plaquettes closest to the link $i$. $d_{i}=\pm 1$ when there is no/one dimer on link $i$ of the reference configuration. Assuming the interaction of visons is weak, this correlation function $C^{vc}_d(r_{i,j},\tau)=\langle D^{vc}_i(0)D^{vc}_j(\tau)\rangle-\langle D^{vc}_i(0)\rangle^2=\langle V_{i_1}(0)V_{i_2}(0)d_iV_{j_1}(\tau)V_{j_2}(\tau)d_j\rangle-\langle V_{i_1}(0)V_{i_2}(0)\rangle^2$ can be computed using Wick's theorem as the convolution of two vison operators,
\begin{equation}
C^{vc}_d(r_{i,j},\tau)=\langle V_{i_1}(0)V_{j_1}(\tau)\rangle \langle V_{i_2}(0)V_{j_2}(\tau)\rangle d_id_j +\langle V_{i_1}(0)V_{j_2}(\tau)\rangle \langle V_{i_2}(0)V_{j_1}(\tau)\rangle d_id_j.
\label{eq:eq2}
\end{equation}
Here, $d_i$ is constant for link $i$ under the gauge choice, and can be taken outside the brackets. The spectrum $C^{vc}_{d}(\mathbf{q},\omega)$, which we refer to as the vison-convolution spectrum, gives rise to the convolution of two vison excitations. It is therefore of great importance to compare it with the dimer spectrum $C_d(\mathbf{q},\omega)$, where the difference will reveal the interaction effects among the visons in different regions of the phase diagram. And we emphasize that although the the bottom of the dimer dispersion has been discussed in the Refs.~\cite{Ralko2006,Ralko2007}, the full numerical calculation of the $C_{d}(\mathbf{q},\omega)$, $C_{v}(\mathbf{q},\omega)$ and $C^{vc}_{d}(\mathbf{q},\omega)$ dynamical correlation functions, both in the frequency and momentum axes, are for the first time being presented here and they provide the well-characterised example of the dynamics of a $Z_2$ spin liquid and a phase transition driven by condensation of fractional excitations.

{\noindent\it Spectra of dimer, vison and vison-convolution.-} In the $Z_2$ QSL phase, the visons are the emergent and fractionalized elementary excitation with no spin and charge quantum numbers~\cite{misguich2008quantum}. As discussed in the introduction, this is an unique advantage of the QDM that single vison spectrum can be measured unambiguously, as usually the vison excitations have to be constructed in mean-field  
as built-in without knowing the unbiased physics~\cite{LiTao2004}, or measured indirectly in lattice models of frustrated magnets~\cite{BFG2002,Isakov2006,YCWang2017QSL,YCWang2018,GYSun2018,YCWangVestigial2020}.

\begin{figure*}[htp]
	\centering
	\includegraphics[width=1\textwidth]{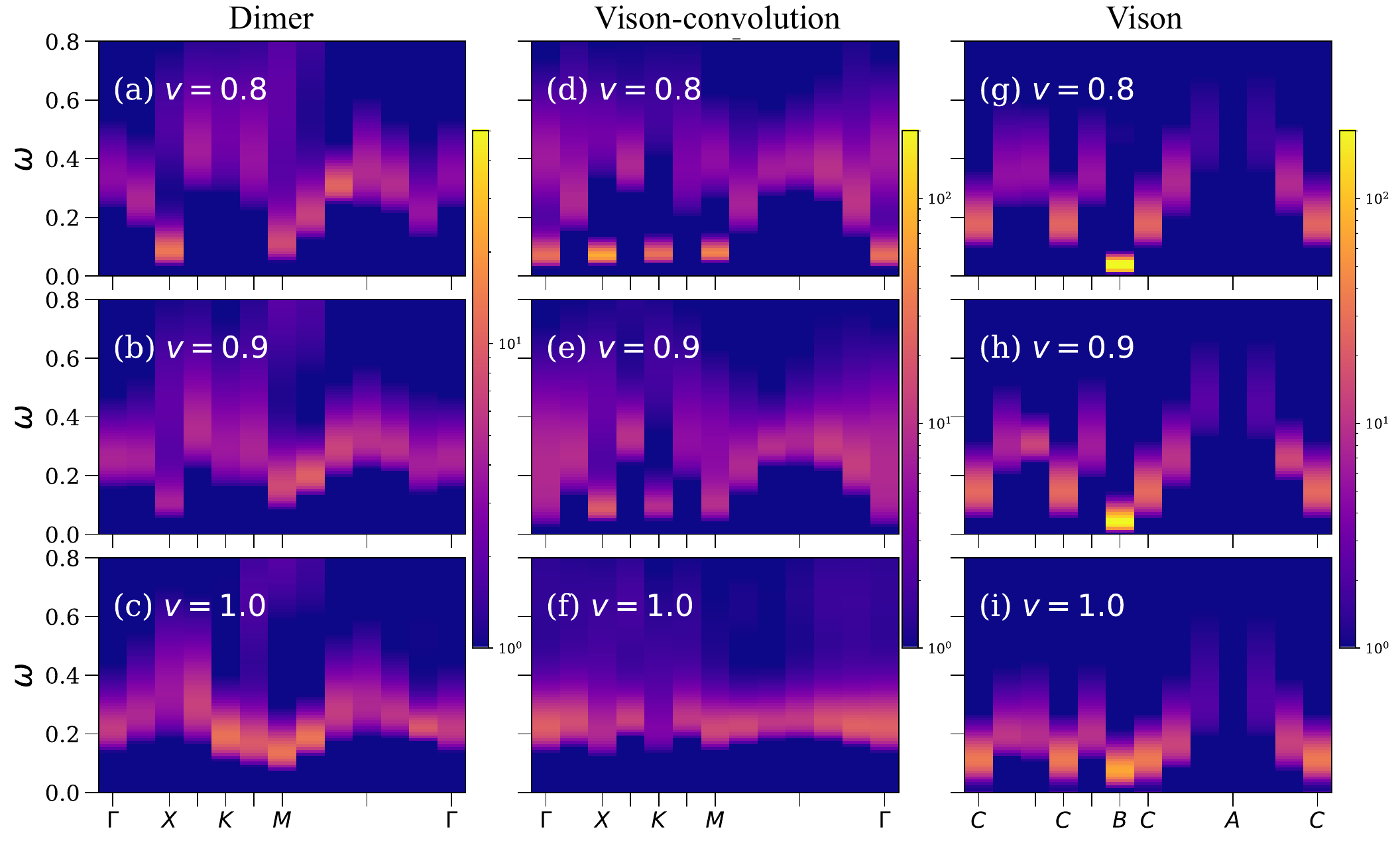}
	\caption{Spectra of dimer (a) (b) (c), vison-convolution (d) (e) (f) and vison (g) (h) (i) correlation functions across the VBS-QSL transition. For (a) (d) (g), (b) (e) (h) and (c) (f) (i), $V=0.8, 0.9, 1$, respectively. The results are obtained from $L=12$ and $\beta=200$ $(T=1/200)$ systems.}
	\label{fig:fig2}
\end{figure*}

We therefore measure the correlation functions of $C_{d}(\mathbf{q},\tau)$, $C_{v}(\mathbf{q},\tau)$ and $C^{vc}_{d}(\mathbf{q},\tau)$ in QMC and then using SAC~\cite{Sandvik2015,Qin2017,GYSun2018} to generate the real frequency spectra $C_{d}(\mathbf{q},\omega)$, $C_{v}(\mathbf{q},\omega)$ and $C^{vc}_{d}(\mathbf{q},\omega)$. These results are presented in Fig.~\ref{fig:fig2}. Inside the $Z_2$ QSL phase with $V=1$, all the spectra are gapped. The vison spectra acquire the smallest gap at the order of $\omega\sim0.1$ at $B$ point of BZ. And the dimer and VC correlations are also gapped with their minimal at $M$ point. It is interesting to notice that the VC spectral gap at $M$ point is higher than the dimer gap at the same point, suggesting that actually visons have a binding energy in forming the dimer correlation and consequently their interaction effect is attractive and gives rise to a bound state with lower energy than the naive convolution. In addition to SAC, we also fit the excitation gaps directly from the imaginary time correlation functions, as shown in Supplemental Material (SM)~\cite{suppl}.

As $V$ is reduced from 1 to 0.9 and 0.8, a QSL-VBS transition is expected at $V_c\sim 0.85$~\cite{Ralko2006,Ralko2007,Ralko2005,Ivanov2004}, and previous works from the gap measurements and field analytical analysis~\cite{MoessnerSondhi2001a} have proposed emergent O(4) symmetry at the transition. But how the entire spectra change across the transition has not been shown due to the lack of access to finite temperature fluctuation effects. With our QMC+SAC scheme, we obseve that the vison gap closes at the $B$ point and the dimer and VC spectrum gap close at $X$ and $M$ points of the BZ (subject to finite size effect of the QMC simulation), as shown in Fig.~\ref{fig:fig2} for $V=0.8$ and $0.9$. The minimal at $\Gamma$ and $K$ of the VC spectra come from the allowed momentum convolution of single vison spectra which has minimal at $B$. Such gap closing process is a manifestation of the symmetry fractionalization mechanism of anyon condensation in $Z_2$ topological order~\cite{QiYang2015a,QiYang2015b,Becker2018,GYSun2018} . That is, since here the $Z_2$ gauge field is odd in nature (see the discussion in SM~\cite{suppl}), the visons carry $\pi$-flux throughout the lattice. As the QSL-VBS critical point is approached, the vison gap will close and the entire vison spectral weight will condensed at a finite momentum point. In a similar manner, the dimer spectra, generated from the vison bound states, will also close at a finite momentum point. This is different from the usual Bose condensation from disorder symmetric state to ordered symmetry-breaking state, where the condensation of the low-lying bosons usually close gap at the $\Gamma$ point. Since in our case the disordered state has intrinsic topological order with elementary excitations (visons) carrying fintie momentum ($\pi$-flux), the condensation gap manifests finite momentum closing. Similar translation symmetry fractionalization process, has also been observed in $\pi$-flux $Z_2$ spin liquid realized in the Kagome lattice model~\cite{QiYang2015a,QiYang2015b,Becker2018,GYSun2018}, which is proposed to be used as a experimental signature of quantum spin liquid in neutron scattering for Kagome antiferromagnet~\cite{Punk2014,Essin2014,JWMei2015}. Also, one sees that at $V=0.9$ and 0.8, there are more higher energy spectral weights in dimer, VC and vison spectra, coming from the enhanced quantum critical fluctuations of the QSL-VBS transition.

{\noindent\it Emergent O(4) symmetry and order parameter of VBS.-} Next we discuss the nature of the QSL-VBS transition and the symmetry breaking pattern of the $\sqrt{12}\times\sqrt{12}$ phase. As explained in the SM~\cite{suppl}, it is expected theoretically~\cite{MoessnerSondhi2001b,Coletta2011Phase} that this transition is driven by the condensation of visons, which is decribed by a four-component order parameter $\{\phi_{i}\}, i=0,1,2,3$ constructed from the Fourier transformation vison configuration at momenta $B$, i.e. $\pm(\frac{\pi}{6},\frac{\pi}{6})$ and $\pm(-\frac{\pi}{6},\frac{5\pi}{6})$ in Fig.~\ref{fig:fig1} (b). The order parameter transforms as a 4D representation under the lattice wallpaper-group symmetries, and the matrix form of group actions are summarized in the SM~\cite{suppl}.

\begin{figure}[htp!]
	\centering
	\includegraphics[width=0.5\columnwidth]{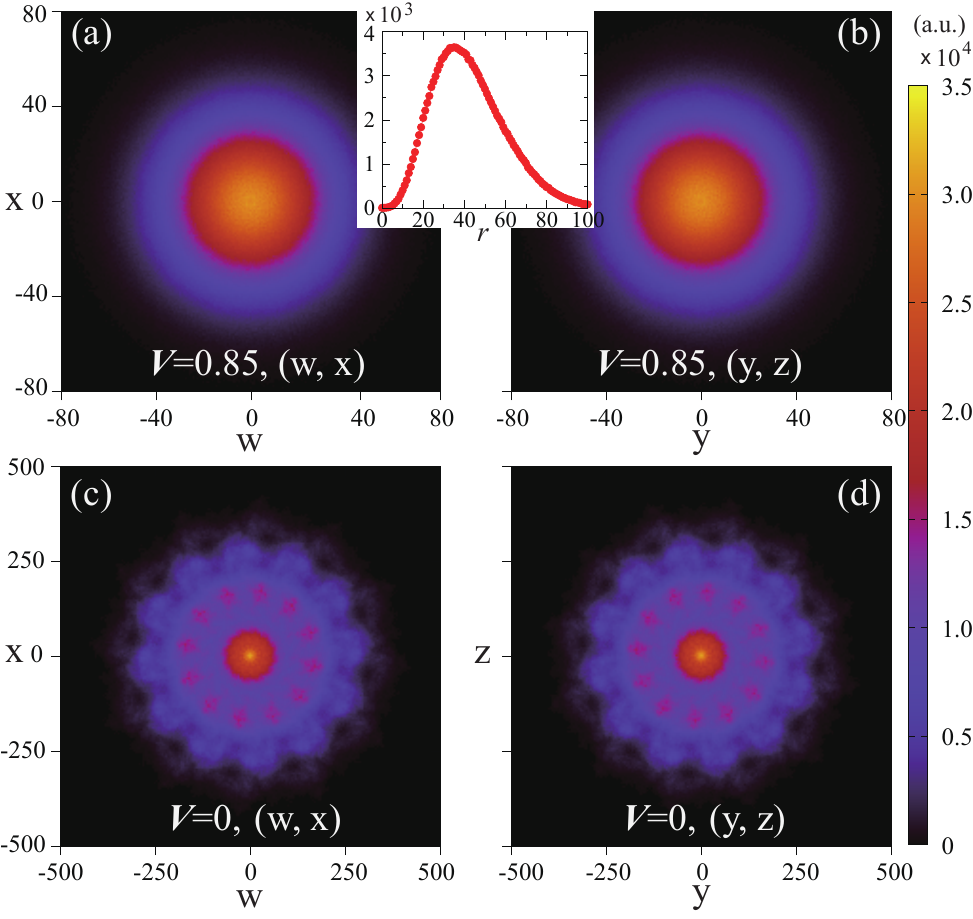}
	\caption{(a), (b) are the projection of the four-dimensional order parameter $(w,x,y,z)$ at the QSL-VBS critical point with $V=0.85$ on the two-dimensional $(w,x)$ plane (a) and on the $(y,z)$ plane (b). (Inset) The density distribution per unit sphere area of the O(4) order parameter modulus at $V=0.85$, such radial dependence reveals that the order parameter indeed form a O(4) sphere. The red line joins the points. (c) and (d) are the projection of the four-dimensional order parameter deep in the $\sqrt{12}\times\sqrt{12}$ VBS phase at $V=0$. The histogram is on the $(w,x)$ plane (c) and the $(y,z)$ plane (d). The data in (a), (b), (c) and (d) are obtained from system size $L=\beta=36$ $(T=1/36)$.}
	\label{fig:fig3}
\end{figure}

In order to numerically confirm that the order parameter $\phi_i$ indeed captures the QSL-VBS transition, we perform a principal component analysis (PCA) on the vison correlation function $C_v$ to extract the condensing mode near the transition.
PCA diagonalizes the $4\times4$ matrix of the momentum-space vison correlation function at the $B$ point, and identifies the eigenvectors with the largest eigenvalues corresponds to the modes representated by the order parameter $\phi_i$. We list the ratio of the first largest eigenvalue over the second at $V=0.5$ to $1$ in Table.~\ref{TB1}. Since the largest eigenvalue always dominate, it shows that the principal component of the VBS structure is indeed the expected $\sqrt{12}\times\sqrt{12}$ order. The theoretical analysis further predicts that, at the QSL-VBS critical point, the transition point acquires an emergent O(4) symmetry, as O(4)-symmetry-breaking terms become irrelevant. In other words, the order parameter lives homogeneously on a four-dimensional sphere~\cite{MoessnerSondhi2001a}.

\begin{table}[htp]
\begin{tabular}{cp{7pt}cp{7pt}cp{7pt}cp{7pt}cp{7pt}cp{7pt}c}
\toprule
  V  && 0.5      && 0.6 && 0.7 && 0.8 && 0.9 && 1 \\
\colrule
  $L_1/L_2$  &&  85.03 &&  82.05 && 76.18 && 68.48 &&  51.36 && 29.71 \\
\botrule
\end{tabular}
\caption{Principal Component Analysis. $L_1/L_2$ means the first largest eigenvalue over the second of the momentum-space vison correlation function matrix at $B$ point. All data are obtained at a $12\times 12$ lattice with $\beta=200$ $(T=1/200)$.}
\label{TB1}
\end{table}


To reveal such emergent O(4) symmetry at the QSL-VBS critical point and its breaking inside the $\sqrt{12}\times\sqrt{12}$ VBS phase. We prepare the order parameter histogram in Fig.~\ref{fig:fig3}. By reorganizing the order parameters into $\phi_0=\phi_2^*=w+ix$ and $\phi_1=\phi_3^*=y+iz$. Since the order parameter is four dimensional and hard to visualize, we draw two-dimensional projected histogram $(w,x)$ and $(y,z)$ of the 4D order parameter near the phase transition point at $V=0.85$ and deep inside the VBS phase at $V=0$. Fig.~\ref{fig:fig3} (a) and (b) are the two independent projections of the 4D $(w,x,y,z)$ space and clearly an emergent O(4) symmetry is present. The Inset shows the modulus distribution of the 4D sphere (with arbitrary unit) which means the order indeed lives homogeneously on a four-dimensional sphere~\cite{YTang2011}. Fig.~\ref{fig:fig3} (c) and (d) are the same analysis inside the VBS phase, and here clearly distinctive points in the two projected phase are present, which are in full consistency with the symmetry analysis in SM~\cite{suppl}, i.e. the $\sqrt{12}\times\sqrt{12}$ VBS breaks the O(4) symmetry.

\section*{Discussion} Via the newly developed sweeping cluster QMC algorithm, supplemented with SAC scheme to obtain the real-frequency data and symmetry analysis of the VBS order parameter, we reveal the excitation spectra in different phases of the triangular lattice QDM, in particular, the single vison excitations inside the $Z_2$ QSL and its condensation towards the $\sqrt{12}\times\sqrt{12}$ VBS with the translational symmetry fractionalization. We found the vison-convolution spectrum is different from the dimer spectrum due to the vison interaction effect, and we also unearth the emergent O(4) symmetry at the QSL-VBS transition and the nature of the $\sqrt{12}\times\sqrt{12}$ VBS with its O(4) order parameter and symmetry breaking. We note that our results not only confirm expectations on triangular lattice QDM by previous works~\cite{MoessnerSondhi2001a,MoessnerSondhi2001b,Ivanov2004,Ralko2005,Ralko2006,Ralko2007}, but more importantly, move forward by directly and reliably characterising the single particle dynamics of fractional excitations using controlled numerics, and demonstrating their condensation towards symmetry-breaking phase. We believe our work provide the well-characterised example of the dynamics of a $Z_2$ spin liquid and opens an avenue for generic solution of the static and dynamic properties of QDMs and other strict constrained systems, such as those in programmable quantum simulators based on Rydberg atom arrays ~\cite{Rhine2021,Ebadi2020,scholl2020} and superconducting qubits~\cite{King2018topology,King2019scalingadv} where geometry frustration and dynamics of quantum Ising models have been proposed	and partially realized.

\section*{methods}
{\noindent\it Sweeping cluster algorithm.-} This is a new quantum Monte Carlo method developed by author which can work well in constrained spin models~\cite{YanZheng2019a,YanZheng2019b,ZhenYan2020}. The key idea of sweeping cluster algorithm is to sweep and update layer by layer along the imaginary time direction, so that the local constraints (gauge field) are recorded by update-lines. Via this way, all the samplings are done in the restricted Hilbert space, i.e. the low-energy space. In this article, we can measure the information of single vison because in a strictly constrained space, the energy gap of other quasi-particles such as spinon, becomes infinite large and thus these quasi-particles does not exist in the restricted Hilbert space. We also note that due to the reduced computional complexity with global updates, the system sizes simulated here is three times larger than those simulated with the projection methods in previous works~\cite{Ralko2005,Ralko2006,Ralko2007}.

{\noindent\it Stochastic analytic continuation.-}
The main idea of this method~\cite{Sandvik1998a,Beach2004,Syljuasen2008} is to obtain the optimal solution of the inverse Laplace transform via sampling depend on importance of goodness. From sweeping cluster method, we can obtain a set of imaginary time correlation functions $G(\tau)$. The real-frequency spectral function and the imaginary time correlation function have the following transformation relationship as $G(\tau)=\frac{1}{\pi}\int_0^\infty d\omega (e^{-\tau\omega}+e^{-(\beta-\tau)\omega}) S(\omega)$. In order to inversely solve this equation, we must fit a better spectral function. Let the spectral function has a general form as $S(\omega)=\sum_{i} a_i \delta(\omega-\omega_i)$. By sampling according to the importance of goodness of fit, we can finally get the spectral function numerically. The reliability of such QMC-SAC scheme has been extensively tested in quantum many-body systems, ranging from 1D Heisenberg chain~\cite{Sandvik2015} compared with Bethe ansatz, 2D Heisenberg model~\cite{Shao2017,CKZhou2020} compared with exact diagonalization, field theoretical analysis and neutron scattering spectra in real square lattice quantum magnets, deconfined quantum critical point~\cite{Shao2017,Ma2018a} and deconfined U(1) spin liquid phase with emergent photon excitations~\cite{CJHuang2018}, $Z_2$ quantum spin liquid model with fractionalized spectra~\cite{GYSun2018,YCWangVestigial2020} compared with anyon condensation theory, to quantum Ising model with direct comparison with neutron scattering and NMR experiments~\cite{Lih2020,ZHu2020}.

\section*{DATA AVAILABILITY}
The data that support the findings of this study are available from the authors upon reasonable request.


\begin{thebibliography}{10}
\expandafter\ifx\csname url\endcsname\relax
  \def\url#1{\texttt{#1}}\fi
\expandafter\ifx\csname urlprefix\endcsname\relax\def\urlprefix{URL }\fi
\providecommand{\bibinfo}[2]{#2}
\providecommand{\eprint}[2][]{\url{#2}}

\bibitem{Wen2019}
\bibinfo{author}{Wen, X.-G.}
\newblock \bibinfo{title}{Choreographed entanglement dances: Topological states
  of quantum matter}.
\newblock \emph{\bibinfo{journal}{Science}} \textbf{\bibinfo{volume}{363}}, eaal3099
  (\bibinfo{year}{2019}).

\bibitem{HanTH12}
\bibinfo{author}{Han, T.~H.} \emph{et~al.}
\newblock \bibinfo{title}{Fractionalized excitations in the spin-liquid state
  of a kagome-lattice antiferromagnet}.
\newblock \emph{\bibinfo{journal}{Nature}} \textbf{\bibinfo{volume}{492}},
  \bibinfo{pages}{406--410} (\bibinfo{year}{2012}).

\bibitem{WeiYuan2017}
\bibinfo{author}{{Wei}, Y.} \emph{et~al.}
\newblock \bibinfo{title}{{Evidence for a $\Z_2$ topological ordered quantum
  spin liquid in a kagome-lattice antiferromagnet}}. {Preprint at https://arxiv.org/abs/1710.02991 ({2017}).}

\bibitem{feng2018claringbullite}
\bibinfo{author}{Feng, Z.} \emph{et~al.}
\newblock \bibinfo{title}{From claringbullite to a new spin liquid candidate
  $Cu_3Zn(OH)_6FCl$}.
\newblock \emph{\bibinfo{journal}{Chinese Physics Letters}}
  \textbf{\bibinfo{volume}{36}}, \bibinfo{pages}{017502}
  (\bibinfo{year}{2018}).

\bibitem{wen2019the}
\bibinfo{author}{Wen, J.~J.} \& \bibinfo{author}{Lee, Y.~S.}
\newblock \bibinfo{title}{The search for the quantum spin liquid in kagome
  antiferromagnets}.
\newblock \emph{\bibinfo{journal}{Chinese Physics Letters}}
  \textbf{\bibinfo{volume}{36}}, \bibinfo{pages}{050101} (\bibinfo{year}{2019}).

\bibitem{YuanWei2020}
\bibinfo{author}{Wei, Y.} \emph{et~al.}
\newblock \bibinfo{title}{Magnetic phase diagram of $Cu_{4-x}Zn_x({OH})_6FBr$ studied by neutron-diffraction and $\mu${SR} techniques}.
\newblock \emph{\bibinfo{journal}{Chinese Physics Letters}}
  \textbf{\bibinfo{volume}{37}}, \bibinfo{pages}{107503}
  (\bibinfo{year}{2020}).

\bibitem{YiZhou2017}
\bibinfo{author}{Zhou, Y.}, \bibinfo{author}{Kanoda, K.} \&
  \bibinfo{author}{Ng, T.-K.}
\newblock \bibinfo{title}{Quantum spin liquid states}.
\newblock \emph{\bibinfo{journal}{Rev. Mod. Phys.}}
  \textbf{\bibinfo{volume}{89}}, \bibinfo{pages}{025003}
  (\bibinfo{year}{2017}).
\newblock

\bibitem{Broholm2020}
\bibinfo{author}{Broholm, C.} \emph{et~al.}
\newblock \bibinfo{title}{Quantum spin liquids}.
\newblock \emph{\bibinfo{journal}{Science}} \textbf{\bibinfo{volume}{367}}, eaay0668
  (\bibinfo{year}{2020}).

\bibitem{Kivelson1987}
\bibinfo{author}{Kivelson, S.~A.}, \bibinfo{author}{Rokhsar, D.~S.} \&
  \bibinfo{author}{Sethna, J.~P.}
\newblock \bibinfo{title}{Topology of the resonating valence-bond state:
  Solitons and high-${T}_{c}$ superconductivity}.
\newblock \emph{\bibinfo{journal}{Phys. Rev. B}} \textbf{\bibinfo{volume}{35}},
  \bibinfo{pages}{8865--8868} (\bibinfo{year}{1987}).

\bibitem{Rokhsar1988}
\bibinfo{author}{Rokhsar, D.~S.} \& \bibinfo{author}{Kivelson, S.~A.}
\newblock \bibinfo{title}{Superconductivity and the quantum hard-core dimer
  gas}.
\newblock \emph{\bibinfo{journal}{Phys. Rev. Lett.}}
  \textbf{\bibinfo{volume}{61}}, \bibinfo{pages}{2376--2379}
  (\bibinfo{year}{1988}).

\bibitem{Baskaran1988}
\bibinfo{author}{Baskaran, G.} \& \bibinfo{author}{Anderson, P.~W.}
\newblock \bibinfo{title}{Gauge theory of high-temperature superconductors and
  strongly correlated Fermi systems}.
\newblock \emph{\bibinfo{journal}{Phys. Rev. B}} \textbf{\bibinfo{volume}{37}},
  \bibinfo{pages}{580--583} (\bibinfo{year}{1988}).

\bibitem{MoessnerSondhi2001a}
\bibinfo{author}{Moessner, R.} \& \bibinfo{author}{Sondhi, S.~L.}
\newblock \bibinfo{title}{Resonating valence bond phase in the triangular
  lattice quantum dimer model}.
\newblock \emph{\bibinfo{journal}{Phys. Rev. Lett.}}
  \textbf{\bibinfo{volume}{86}}, \bibinfo{pages}{1881--1884}
  (\bibinfo{year}{2001}).

\bibitem{MoessnerSondhi2001b}
\bibinfo{author}{Moessner, R.} \& \bibinfo{author}{Sondhi, S.~L.}
\newblock \bibinfo{title}{Ising models of quantum frustration}.
\newblock \emph{\bibinfo{journal}{Phys. Rev. B}} \textbf{\bibinfo{volume}{63}},
  \bibinfo{pages}{224401} (\bibinfo{year}{2001}).

\bibitem{furukawa2007topological}
\bibinfo{author}{Furukawa, S.} \& \bibinfo{author}{Misguich, G.}
\newblock \bibinfo{title}{Topological entanglement entropy in the quantum dimer
  model on the triangular lattice}.
\newblock \emph{\bibinfo{journal}{Phys. Rev. B}} \textbf{\bibinfo{volume}{75}},
  \bibinfo{pages}{214407} (\bibinfo{year}{2007}).

\bibitem{Ivanov2004}
\bibinfo{author}{Ivanov, D.~A.}
\newblock \bibinfo{title}{Vortexlike elementary excitations in the
  Rokhsar-Kivelson dimer model on the triangular lattice}.
\newblock \emph{\bibinfo{journal}{Phys. Rev. B}} \textbf{\bibinfo{volume}{70}},
  \bibinfo{pages}{094430} (\bibinfo{year}{2004}).

\bibitem{Ralko2005}
\bibinfo{author}{Ralko, A.}, \bibinfo{author}{Ferrero, M.},
  \bibinfo{author}{Becca, F.}, \bibinfo{author}{Ivanov, D.} \&
  \bibinfo{author}{Mila, F.}
\newblock \bibinfo{title}{Zero-temperature properties of the quantum dimer
  model on the triangular lattice}.
\newblock \emph{\bibinfo{journal}{Phys. Rev. B}} \textbf{\bibinfo{volume}{71}},
  \bibinfo{pages}{224109} (\bibinfo{year}{2005}).

\bibitem{Ralko2006}
\bibinfo{author}{Ralko, A.}, \bibinfo{author}{Ferrero, M.},
  \bibinfo{author}{Becca, F.}, \bibinfo{author}{Ivanov, D.} \&
  \bibinfo{author}{Mila, F.}
\newblock \bibinfo{title}{Dynamics of the quantum dimer model on the triangular
  lattice: Soft modes and local resonating valence-bond correlations}.
\newblock \emph{\bibinfo{journal}{Phys. Rev. B}} \textbf{\bibinfo{volume}{74}},
  \bibinfo{pages}{134301} (\bibinfo{year}{2006}).

\bibitem{Ralko2007}
\bibinfo{author}{Ralko, A.}, \bibinfo{author}{Ferrero, M.},
  \bibinfo{author}{Becca, F.}, \bibinfo{author}{Ivanov, D.} \&
  \bibinfo{author}{Mila, F.}
\newblock \bibinfo{title}{Crystallization of the resonating valence bond liquid
  as vortex condensation}.
\newblock \emph{\bibinfo{journal}{Phys. Rev. B}} \textbf{\bibinfo{volume}{76}},
  \bibinfo{pages}{140404} (\bibinfo{year}{2007}).

\bibitem{YanZheng2019a}
\bibinfo{author}{Yan, Z.} \emph{et~al.}
\newblock \bibinfo{title}{Sweeping cluster algorithm for quantum spin systems
  with strong geometric restrictions}.
\newblock \emph{\bibinfo{journal}{Phys. Rev. B}} \textbf{\bibinfo{volume}{99}},
  \bibinfo{pages}{165135} (\bibinfo{year}{2019}).

\bibitem{YanZheng2019b}
\bibinfo{author}{{Yan}, Z.} \emph{et~al.}
\newblock \bibinfo{title}{{Widely existing mixed phase structure of quantum
  dimer model on square lattice}}. {Preprint at https://arxiv.org/abs/1911.05433 (2019).}

\bibitem{ZhenYan2020}
\bibinfo{author}{{Yan}, Z.}
\newblock \bibinfo{title}{{Improved sweeping cluster algorithm for quantum
  dimer model}}. {Preprint at https://arxiv.org/abs/2011.08457 (2020).}

\bibitem{Laeuchli2008}
\bibinfo{author}{Läuchli, A.~M.}, \bibinfo{author}{Capponi, S.} \&
  \bibinfo{author}{Assaad, F.~F.}
\newblock \bibinfo{title}{Dynamical dimer correlations at bipartite and
  non-bipartite Rokhsar{\textendash}Kivelson points}.
\newblock \emph{\bibinfo{journal}{Journal of Statistical Mechanics: Theory and
  Experiment}} \textbf{\bibinfo{volume}{2008}}, \bibinfo{pages}{P01010}
  (\bibinfo{year}{2008}).

\bibitem{Henley2004}
\bibinfo{author}{Henley, C.~L.}
\newblock \bibinfo{title}{From classical to quantum dynamics at
  Rokhsar{\textendash}Kivelson points}.
\newblock \emph{\bibinfo{journal}{Journal of Physics: Condensed Matter}}
  \textbf{\bibinfo{volume}{16}}, \bibinfo{pages}{S891--S898}
  (\bibinfo{year}{2004}).

\bibitem{OFS2005}
\bibinfo{author}{Sylju{\aa}sen, O.~F.}
\newblock \bibinfo{title}{Continuous-time diffusion Monte Carlo method applied
  to the quantum dimer model}.
\newblock \emph{\bibinfo{journal}{Physical Review B}}
  \textbf{\bibinfo{volume}{71}}, \bibinfo{pages}{020401}
  (\bibinfo{year}{2005}).

\bibitem{OFS2006}
\bibinfo{author}{Sylju{\aa}sen, O.~F.}
\newblock \bibinfo{title}{Plaquette phase of the square-lattice quantum dimer
  model: Quantum Monte Carlo calculations}.
\newblock \emph{\bibinfo{journal}{Physical Review B}}
  \textbf{\bibinfo{volume}{73}}, \bibinfo{pages}{245105}
  (\bibinfo{year}{2006}).

\bibitem{OFS2005walk}
\bibinfo{author}{Sylju{\aa}sen, O.~F.}
\newblock \bibinfo{title}{Random walks near Rokhsar--Kivelson points}.
\newblock \emph{\bibinfo{journal}{International Journal of Modern Physics B}}
  \textbf{\bibinfo{volume}{19}}, \bibinfo{pages}{1973--1993}
  (\bibinfo{year}{2005}).

\bibitem{OFS2002}
\bibinfo{author}{Syljuasen, O.~F.} \& \bibinfo{author}{Sandvik, A.~W.}
\newblock \bibinfo{title}{Quantum Monte Carlo with directed loops.}
\newblock \emph{\bibinfo{journal}{Physical Review E}}
  \textbf{\bibinfo{volume}{66}}, \bibinfo{pages}{046701}
  (\bibinfo{year}{2002}).

\bibitem{Alet2005a}
\bibinfo{author}{Alet, F.}, \bibinfo{author}{Wessel, S.} \&
  \bibinfo{author}{Troyer, M.}
\newblock \bibinfo{title}{Generalized directed loop method for quantum monte
  carlo simulations}.
\newblock \emph{\bibinfo{journal}{Phys. Rev. E}} \textbf{\bibinfo{volume}{71}},
  \bibinfo{pages}{036706} (\bibinfo{year}{2005}).

\bibitem{Alet2005b}
\bibinfo{author}{Alet, F.} \emph{et~al.}
\newblock \bibinfo{title}{Interacting classical dimers on the square lattice}.
\newblock \emph{\bibinfo{journal}{Phys. Rev. Lett.}}
  \textbf{\bibinfo{volume}{94}}, \bibinfo{pages}{235702}
  (\bibinfo{year}{2005}).

\bibitem{Sandvik1998a}
\bibinfo{author}{Sandvik, A.~W.}
\newblock \bibinfo{title}{Stochastic method for analytic continuation of
  quantum Monte Carlo data}.
\newblock \emph{\bibinfo{journal}{Phys. Rev. B}} \textbf{\bibinfo{volume}{57}},
  \bibinfo{pages}{10287--10290} (\bibinfo{year}{1998}).

\bibitem{Beach2004}
\bibinfo{author}{{Beach}, K.~S.~D.}
\newblock \bibinfo{title}{{Identifying the maximum entropy method as a special
  limit of stochastic analytic continuation}}. {Preprint at https://arxiv.org/abs/cond-mat/0403055 (2004).}

\bibitem{Syljuasen2008}
\bibinfo{author}{Sylju\aa{}sen, O.~F.}
\newblock \bibinfo{title}{Using the average spectrum method to extract dynamics
  from quantum Monte Carlo simulations}.
\newblock \emph{\bibinfo{journal}{Phys. Rev. B}} \textbf{\bibinfo{volume}{78}},
  \bibinfo{pages}{174429} (\bibinfo{year}{2008}).

\bibitem{Sandvik2015}
\bibinfo{author}{Sandvik, A.~W.}
\newblock \bibinfo{title}{Constrained sampling method for analytic
  continuation}.
\newblock \emph{\bibinfo{journal}{Phys. Rev. E}} \textbf{\bibinfo{volume}{94}},
  \bibinfo{pages}{063308} (\bibinfo{year}{2016}).

\bibitem{Qin2017}
\bibinfo{author}{Qin, Y.~Q.}, \bibinfo{author}{Normand, B.},
  \bibinfo{author}{Sandvik, A.~W.} \& \bibinfo{author}{Meng, Z.~Y.}
\newblock \bibinfo{title}{Amplitude mode in three-dimensional dimerized
  antiferromagnets}.
\newblock \emph{\bibinfo{journal}{Phys. Rev. Lett.}}
  \textbf{\bibinfo{volume}{118}}, \bibinfo{pages}{147207}
  (\bibinfo{year}{2017}).

\bibitem{GYSun2018}
\bibinfo{author}{Sun, G.-Y.} \emph{et~al.}
\newblock \bibinfo{title}{Dynamical signature of symmetry fractionalization in
  frustrated magnets}.
\newblock \emph{\bibinfo{journal}{Phys. Rev. Lett.}}
  \textbf{\bibinfo{volume}{121}}, \bibinfo{pages}{077201}
  (\bibinfo{year}{2018}).

\bibitem{Shao2017}
\bibinfo{author}{Shao, H.} \emph{et~al.}
\newblock \bibinfo{title}{Nearly deconfined spinon excitations in the
  square-lattice spin-$1/2$ Heisenberg antiferromagnet}.
\newblock \emph{\bibinfo{journal}{Phys. Rev. X}} \textbf{\bibinfo{volume}{7}},
  \bibinfo{pages}{041072} (\bibinfo{year}{2017}).

\bibitem{Ma2018a}
\bibinfo{author}{Ma, N.} \emph{et~al.}
\newblock \bibinfo{title}{Dynamical signature of fractionalization at a
  deconfined quantum critical point}.
\newblock \emph{\bibinfo{journal}{Phys. Rev. B}} \textbf{\bibinfo{volume}{98}},
  \bibinfo{pages}{174421} (\bibinfo{year}{2018}).

\bibitem{CKZhou2020}
\bibinfo{author}{{Zhou}, C.}, \bibinfo{author}{{Yan}, Z.},
  \bibinfo{author}{{Sun}, K.}, \bibinfo{author}{{Starykh}, O.~A.} \&
  \bibinfo{author}{{Meng}, Z.~Y.}
\newblock \bibinfo{title}{{Amplitude mode in quantum magnets via dimensional
  crossover}}. {Preprint at https://arxiv.org/abs/2007.12715 (2020).}

\bibitem{YCWangVestigial2020}
\bibinfo{author}{Wang, Y.-C.}, \bibinfo{author}{Yan, Z.},
  \bibinfo{author}{Wang, C.}, \bibinfo{author}{Qi, Y.} \&
  \bibinfo{author}{Meng, Z.~Y.}
\newblock \bibinfo{title}{Vestigial anyon condensation in kagome quantum spin
  liquids}.
\newblock \emph{\bibinfo{journal}{Phys. Rev. B}}
  \textbf{\bibinfo{volume}{103}}, \bibinfo{pages}{014408}
  (\bibinfo{year}{2021}).

\bibitem{Lih2020}
\bibinfo{author}{Li, H.} \emph{et~al.}
\newblock \bibinfo{title}{{Kosterlitz-Thouless} melting of magnetic order in
  the triangular quantum {Ising} material {TmMgGaO$_4$}}.
\newblock \emph{\bibinfo{journal}{Nat. Commun.}} \textbf{\bibinfo{volume}{11}},
  \bibinfo{pages}{1111} (\bibinfo{year}{2020}).

\bibitem{ZHu2020}
\bibinfo{author}{Hu, Z.} \emph{et~al.}
\newblock \bibinfo{title}{Evidence of the Berezinskii-Kosterlitz-Thouless phase
  in a frustrated magnet}.
\newblock \emph{\bibinfo{journal}{Nature Communications}}
  \textbf{\bibinfo{volume}{11}}, \bibinfo{pages}{5631} (\bibinfo{year}{2020}).

\bibitem{misguich2008quantum}
\bibinfo{author}{Misguich, G.} \& \bibinfo{author}{Mila, F.}
\newblock \bibinfo{title}{Quantum dimer model on the triangular lattice:
  Semiclassical and variational approaches to vison dispersion and
  condensation}.
\newblock \emph{\bibinfo{journal}{Phys. Rev. B}} \textbf{\bibinfo{volume}{77}},
  \bibinfo{pages}{134421} (\bibinfo{year}{2008}).

\bibitem{LiTao2004}
\bibinfo{author}{Li, T.} \& \bibinfo{author}{Yang, H.-Y.}
\newblock \bibinfo{title}{Topological order in Gutzwiller-projected wave
  functions for quantum antiferromagnets}.
\newblock \emph{\bibinfo{journal}{Phys. Rev. B}} \textbf{\bibinfo{volume}{75}},
  \bibinfo{pages}{172502} (\bibinfo{year}{2007}).

\bibitem{BFG2002}
\bibinfo{author}{Balents, L.}, \bibinfo{author}{Fisher, M. P.~A.} \&
  \bibinfo{author}{Girvin, S.~M.}
\newblock \bibinfo{title}{Fractionalization in an easy-axis kagome
  antiferromagnet}.
\newblock \emph{\bibinfo{journal}{Phys. Rev. B}} \textbf{\bibinfo{volume}{65}},
  \bibinfo{pages}{224412} (\bibinfo{year}{2002}).

\bibitem{Isakov2006}
\bibinfo{author}{Isakov, S.~V.}, \bibinfo{author}{Kim, Y.~B.} \&
  \bibinfo{author}{Paramekanti, A.}
\newblock \bibinfo{title}{Spin-liquid phase in a spin-$1/2$ quantum magnet on
  the kagome lattice}.
\newblock \emph{\bibinfo{journal}{Phys. Rev. Lett.}}
  \textbf{\bibinfo{volume}{97}}, \bibinfo{pages}{207204}
  (\bibinfo{year}{2006}).

\bibitem{YCWang2017QSL}
\bibinfo{author}{{Wang}, Y.-C.}, \bibinfo{author}{{Fang}, C.},
  \bibinfo{author}{{Cheng}, M.}, \bibinfo{author}{{Qi}, Y.} \&
  \bibinfo{author}{{Meng}, Z.~Y.}
\newblock \bibinfo{title}{{Topological spin liquid with symmetry-protected edge
  states}}. {Preprint at https://arxiv.org/abs/1701.01552 (2017).}

\bibitem{YCWang2018}
\bibinfo{author}{Wang, Y.-C.}, \bibinfo{author}{Zhang, X.-F.},
  \bibinfo{author}{Pollmann, F.}, \bibinfo{author}{Cheng, M.} \&
  \bibinfo{author}{Meng, Z.~Y.}
\newblock \bibinfo{title}{Quantum spin liquid with even Ising gauge field
  structure on kagome lattice}.
\newblock \emph{\bibinfo{journal}{Phys. Rev. Lett.}}
  \textbf{\bibinfo{volume}{121}}, \bibinfo{pages}{057202}
  (\bibinfo{year}{2018}).

\bibitem{suppl}
\emph{\bibinfo{journal}{Derivations of the vison operators, the O(4) order
  parameter of the VBS phase and its symmetry transformation, and examples of
  excitation gaps obtained directly from fitting the imaginary time decay of
  the correlation functions}} .

\bibitem{QiYang2015a}
\bibinfo{author}{Qi, Y.} \& \bibinfo{author}{Fu, L.}
\newblock \bibinfo{title}{Anomalous crystal symmetry fractionalization on the
  surface of topological crystalline insulators}.
\newblock \emph{\bibinfo{journal}{Phys. Rev. Lett.}}
  \textbf{\bibinfo{volume}{115}}, \bibinfo{pages}{236801}
  (\bibinfo{year}{2015}).

\bibitem{QiYang2015b}
\bibinfo{author}{{Qi}, Y.}, \bibinfo{author}{{Cheng}, M.} \&
  \bibinfo{author}{{Fang}, C.}
\newblock \bibinfo{title}{{Symmetry fractionalization of visons in $\mathbb
  Z_2$ spin liquids}}. {Preprint at https://arxiv.org/abs/1509.02927 (2015).}

\bibitem{Becker2018}
\bibinfo{author}{Becker, J.} \& \bibinfo{author}{Wessel, S.}
\newblock \bibinfo{title}{Diagnosing fractionalization from the spin dynamics
  of ${Z}_{2}$ spin liquids on the kagome lattice by quantum Monte Carlo
  simulations}.
\newblock \emph{\bibinfo{journal}{Phys. Rev. Lett.}}
  \textbf{\bibinfo{volume}{121}}, \bibinfo{pages}{077202}
  (\bibinfo{year}{2018}).

\bibitem{Punk2014}
\bibinfo{author}{Punk, M.}, \bibinfo{author}{Chowdhury, D.} \&
  \bibinfo{author}{Sachdev, S.}
\newblock \bibinfo{title}{Topological excitations and the dynamic structure
  factor of spin liquids on the kagome lattice}.
\newblock \emph{\bibinfo{journal}{Nature Physics}}
  \textbf{\bibinfo{volume}{10}}, \bibinfo{pages}{289 -- 293}
  (\bibinfo{year}{2014}).

\bibitem{Essin2014}
\bibinfo{author}{Essin, A.~M.} \& \bibinfo{author}{Hermele, M.}
\newblock \bibinfo{title}{Spectroscopic signatures of crystal momentum
  fractionalization}.
\newblock \emph{\bibinfo{journal}{Phys. Rev. B}} \textbf{\bibinfo{volume}{90}},
  \bibinfo{pages}{121102} (\bibinfo{year}{2014}).

\bibitem{JWMei2015}
\bibinfo{author}{{Mei}, J.-W.} \& \bibinfo{author}{{Wen}, X.-G.}
\newblock \bibinfo{title}{{Fractionalized spin-wave continuum in spin liquid
  states on the kagome lattice}}. {Preprint at https://arxiv.org/abs/1507.03007 (2015).}

\bibitem{Coletta2011Phase}
\bibinfo{author}{Coletta, T.}, \bibinfo{author}{Picon, J.-D.},
  \bibinfo{author}{Korshunov, S.~E.} \& \bibinfo{author}{Mila, F.}
\newblock \bibinfo{title}{Phase diagram of the fully frustrated
  transverse-field Ising model on the honeycomb lattice}.
\newblock \emph{\bibinfo{journal}{Phys. Rev. B}} \textbf{\bibinfo{volume}{83}},
  \bibinfo{pages}{054402} (\bibinfo{year}{2011}).

\bibitem{YTang2011}
\bibinfo{author}{Tang, Y.}, \bibinfo{author}{Sandvik, A.~W.} \&
  \bibinfo{author}{Henley, C.~L.}
\newblock \bibinfo{title}{Properties of resonating-valence-bond spin liquids
  and critical dimer models}.
\newblock \emph{\bibinfo{journal}{Phys. Rev. B}} \textbf{\bibinfo{volume}{84}},
  \bibinfo{pages}{174427} (\bibinfo{year}{2011}).

\bibitem{Rhine2021}
\bibinfo{author}{Samajdar, R.}, \bibinfo{author}{Ho, W.~W.},
  \bibinfo{author}{Pichler, H.}, \bibinfo{author}{Lukin, M.~D.} \&
  \bibinfo{author}{Sachdev, S.}
\newblock \bibinfo{title}{Quantum phases of {Rydberg} atoms on a kagome
  lattice}.
\newblock \emph{\bibinfo{journal}{Proceedings of the National Academy of
  Sciences}} \textbf{\bibinfo{volume}{118}}, 021034 (\bibinfo{year}{2021}).

\bibitem{Ebadi2020}
\bibinfo{author}{{Ebadi}, S.} \emph{et~al.}
\newblock \bibinfo{title}{{Quantum phases of matter on a 256-atom programmable
  quantum simulator}}. {Preprint at https://arxiv.org/abs/2012.12281 (2020).}

\bibitem{scholl2020}
\bibinfo{author}{{Scholl}, P.} \emph{et~al.}
\newblock \bibinfo{title}{{Programmable quantum simulation of 2D
  antiferromagnets with hundreds of Rydberg atoms}}. {Preprint at https://arxiv.org/abs/2012.12268 (2020).}

\bibitem{King2018topology}
\bibinfo{author}{King, A.~D.} \emph{et~al.}
\newblock \bibinfo{title}{{Observation of topological phenomena in a
  programmable lattice of 1,800 qubits}}.
\newblock \emph{\bibinfo{journal}{Nature}} \textbf{\bibinfo{volume}{560}},
  \bibinfo{pages}{456--460} (\bibinfo{year}{2018}).

\bibitem{King2019scalingadv}
\bibinfo{author}{{King}, A.~D.} \emph{et~al.}
\newblock \bibinfo{title}{{Scaling advantage in quantum simulation of
  geometrically frustrated magnets}}. {Preprint at https://arxiv.org/abs/1911.03446 (2019).}

\bibitem{CJHuang2018}
\bibinfo{author}{Huang, C.-J.}, \bibinfo{author}{Deng, Y.},
  \bibinfo{author}{Wan, Y.} \& \bibinfo{author}{Meng, Z.~Y.}
\newblock \bibinfo{title}{Dynamics of topological excitations in a model
  quantum spin ice}.
\newblock \emph{\bibinfo{journal}{Phys. Rev. Lett.}}
  \textbf{\bibinfo{volume}{120}}, \bibinfo{pages}{167202}
  (\bibinfo{year}{2018}).

\end{thebibliography}

\begin{thebibliography}{2}%
\makeatletter
\providecommand \@ifxundefined [1]{%
 \@ifx{#1\undefined}
}%
\providecommand \@ifnum [1]{%
 \ifnum #1\expandafter \@firstoftwo
 \else \expandafter \@secondoftwo
 \fi
}%
\providecommand \@ifx [1]{%
 \ifx #1\expandafter \@firstoftwo
 \else \expandafter \@secondoftwo
 \fi
}%
\providecommand \natexlab [1]{#1}%
\providecommand \enquote  [1]{``#1''}%
\providecommand \bibnamefont  [1]{#1}%
\providecommand \bibfnamefont [1]{#1}%
\providecommand \citenamefont [1]{#1}%
\providecommand \href@noop [0]{\@secondoftwo}%
\providecommand \href [0]{\begingroup \@sanitize@url \@href}%
\providecommand \@href[1]{\@@startlink{#1}\@@href}%
\providecommand \@@href[1]{\endgroup#1\@@endlink}%
\providecommand \@sanitize@url [0]{\catcode `\\12\catcode `\$12\catcode
  `\&12\catcode `\#12\catcode `\^12\catcode `\_12\catcode `\%12\relax}%
\providecommand \@@startlink[1]{}%
\providecommand \@@endlink[0]{}%
\providecommand \url  [0]{\begingroup\@sanitize@url \@url }%
\providecommand \@url [1]{\endgroup\@href {#1}{\urlprefix }}%
\providecommand \urlprefix  [0]{URL }%
\providecommand \Eprint [0]{\href }%
\providecommand \doibase [0]{http://dx.doi.org/}%
\providecommand \selectlanguage [0]{\@gobble}%
\providecommand \bibinfo  [0]{\@secondoftwo}%
\providecommand \bibfield  [0]{\@secondoftwo}%
\providecommand \translation [1]{[#1]}%
\providecommand \BibitemOpen [0]{}%
\providecommand \bibitemStop [0]{}%
\providecommand \bibitemNoStop [0]{.\EOS\space}%
\providecommand \EOS [0]{\spacefactor3000\relax}%
\providecommand \BibitemShut  [1]{\csname bibitem#1\endcsname}%
\let\auto@bib@innerbib\@empty
\bibitem [{\citenamefont {Moessner}\ and\ \citenamefont
  {Sondhi}(2001)}]{MoessnerSondhi2001b}%
  \BibitemOpen
  \bibfield  {author} {\bibinfo {author} {\bibfnamefont {R.}~\bibnamefont
  {Moessner}}\ and\ \bibinfo {author} {\bibfnamefont {S.~L.}\ \bibnamefont
  {Sondhi}},\ }\href {\doibase 10.1103/PhysRevB.63.224401} {\bibfield
  {journal} {\bibinfo  {journal} {Phys. Rev. B}\ }\textbf {\bibinfo {volume}
  {63}},\ \bibinfo {pages} {224401} (\bibinfo {year} {2001})}\BibitemShut
  {NoStop}%
\bibitem [{\citenamefont {Ralko}\ \emph {et~al.}(2005)\citenamefont {Ralko},
  \citenamefont {Ferrero}, \citenamefont {Becca}, \citenamefont {Ivanov},\ and\
  \citenamefont {Mila}}]{Ralko2005}%
  \BibitemOpen
  \bibfield  {author} {\bibinfo {author} {\bibfnamefont {A.}~\bibnamefont
  {Ralko}}, \bibinfo {author} {\bibfnamefont {M.}~\bibnamefont {Ferrero}},
  \bibinfo {author} {\bibfnamefont {F.}~\bibnamefont {Becca}}, \bibinfo
  {author} {\bibfnamefont {D.}~\bibnamefont {Ivanov}}, \ and\ \bibinfo {author}
  {\bibfnamefont {F.}~\bibnamefont {Mila}},\ }\href {\doibase
  10.1103/PhysRevB.71.224109} {\bibfield  {journal} {\bibinfo  {journal} {Phys.
  Rev. B}\ }\textbf {\bibinfo {volume} {71}},\ \bibinfo {pages} {224109}
  (\bibinfo {year} {2005})}\BibitemShut {NoStop}%
\end{thebibliography}

\section*{Acknowledgement}
We thank Anders W. Sandvik, Andreas L\"auchli, Ying-Jer Kao, Jonathan D'Emidio, Zheng Zhou and Yuan Wan for insightful discussions. ZY and ZYM acknowledge the support from the RGC of Hong Kong SAR of China
(Grant Nos. 17303019 and 17301420), MOST through the National Key Research and Development Program (Grant
No. 2016YFA0300502) and the Strategic Priority Research Program of the Chinese Academy of Sciences (Grant No.
XDB33000000). YQ acknowledges supports from MOST under Grant No. 2015CB921700, and from NSFC under Grant No. 11874115. YCW acknowledges the supports from the NSFC under Grant No. 11804383, the NSF of Jiangsu Province under Grant No. BK20180637, and the Fundamental Research Funds for the Central Universities under Grant No. 2018QNA39. We thank the Computational Initiative at the Faculty of Science and the Information Technology Service at the University of Hong Kong and the Tianhe-1A, Tianhe-2 and Tianhe 3 prototype platforms at the National Supercomputer Centers in Tianjin and Guangzhou for their technical support and generous allocation of CPU time.

\section*{AUTHOR CONTRIBUTIONS}
Y.Q. and Z.Y.M. initiated the work. Z.Y. and Y.C.W. performed the computational simulations. All authors contributed to the analysis of the results. Y.Q. and Z.Y.M. supervised the project.

\section*{COMPETING INTERESTS}
The authors declare no competing interests.

\setcounter{page}{1}
\setcounter{equation}{0}
\setcounter{figure}{0}
\renewcommand{\theequation}{S\arabic{equation}}
\renewcommand{\thefigure}{S\arabic{figure}}

\section*{supplemental material}
In this supplemental material, we analyze the symmetry of the vison order parameter, associated with the phase transition between the $\mathbb Z_2$ QSL and the $\sqrt{12}\times\sqrt{12}$ VBS phases in the QDM on triangular lattices.
Intuitively, such a phase transition is driven by the condensation of visons, which belong to a type of fractional excitations in the $\mathbb Z_2$ QSL.
The correlation function of visons can be described using the vison-string operator,
defined on a string $L:I\rightarrow J$ connecting two sites of the dual lattice:
\begin{equation}
	\label{eq:vstr}
	V_{L} = \prod_{\langle ij\rangle\in L}(-1)^{n_{ij}},
\end{equation}
where the product goes over each bond $\langle ij\rangle$ cut by the string $L$ on the dual lattice, as shown in Fig.~S1(a).

\section{Vison operators}
\label{sec:vop}

To analyze the vison condensation, it is convenient to map the quantum dimer model to a fully-frustrated transverse-field Ising model (FFTFIM) on the dual lattice, which is a honeycomb lattice~\cite{MoessnerSondhi2001b}.
The mapping is illustrated in Fig.~S1(b):
Two spins on nearest-neighbor sites are frustrated (unfrustrated) if there is a dimer (no dimer) on the bond separating them, respectively.
In other words, the spins and dimer occupation satisfies the following relation,
\begin{equation}
	\label{eq:Jss=n}
	J_{IJ}s^z_Is^z_J = (-1)^{n_{ij}}.
\end{equation}
Here, the bond $\langle ij\rangle$ separates the two dual-lattice sites $I$ and $J$.
$J_{IJ}$ is a fully-frustrated Ising coupling: there is exactly one antiferromagnetic coupling $J_{IJ}=-1$ on each hexagon of the dual honeycomb lattice.
The pattern of $J_{IJ}=\pm1$ is a choice of gauge.
Without losing generality, we follow Ref.~\cite{MoessnerSondhi2001b} but choose another gauge shown in Fig.~S1(c).
This mapping is neither surjective nor injective.
On one hand, it maps each dimer configuration to two spin configurations that are exactly opposite to each other, because the two spin configurations $(s_I^z,s_J^z)$ and $(-s_I^z,-s_J^z)$ give the same $n_{ij}$ in Eq.~\eqref{eq:Jss=n}.
On the other hand, dimer configurations satisfying the one-dimer-per-site rule only maps to a subset of spin configurations which minimize the energy of the fully-frustrated Ising model (FFIM).
In this way, the original quantum dimer model can be mapped to the following FFTFIM,
\begin{equation}
	\label{eq:fftfim}
	H = -\sum_{\langle IJ\rangle}J_{IJ}s_I^zs_J^z
	-\Gamma\sum_Is_I^x,
\end{equation}
in the small-$\Gamma$ limit, $\Gamma\ll |J_{IJ}|$.
On the level of Hamiltonians, the mapping between the QDM and the FFTFIM is only approximate and only valid at one parameter point.
However, on the level of quantum states, the mapping is always valid between dimer configurations and spin configurations in the low-energy sector.
Therefore, the map can be used to construct order parameters and analyze the phase transition of vison condensation at different parameters across the phasel transition.

\begin{figure}[htp]
\includegraphics[width=1\columnwidth]{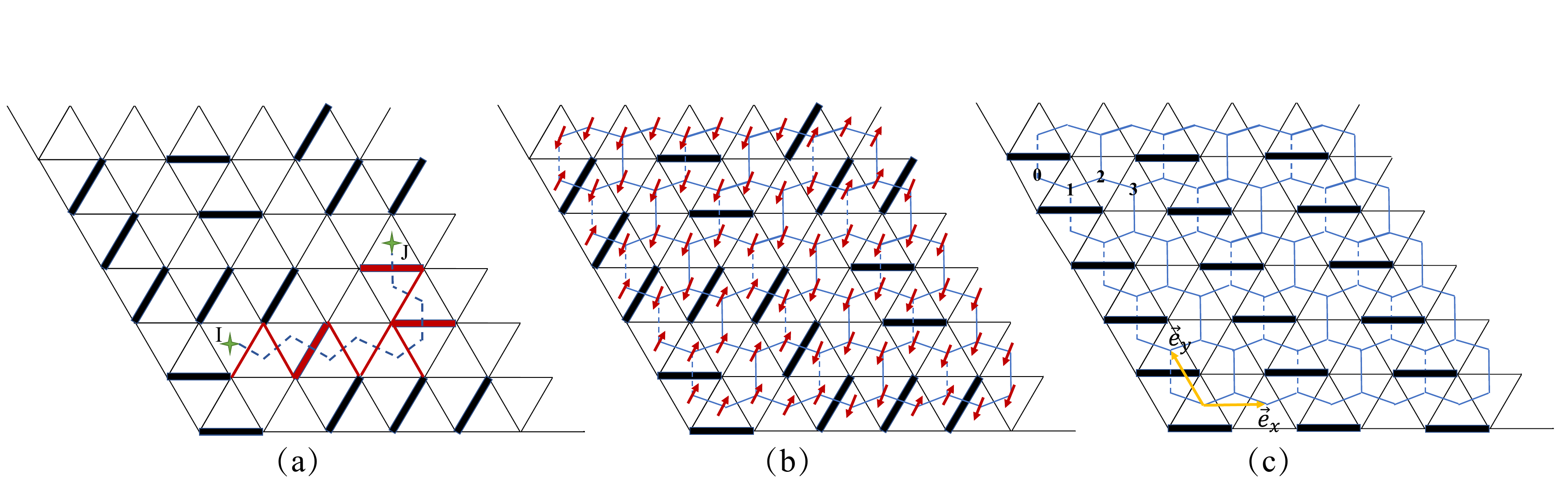}
\caption{All the data in this article are simulated on triangular lattice with periodic boundary condition, (a), (b) and (c) show a $6\times 6$ lattice example, which is the type-A cluster in Ref.\cite{Ralko2005}. (a) Vison-string operator and cut links. (b) Map the quantum dimer model to a fully-frustrated transverse-field Ising model (FFTFIM) on the dual lattice. (c) The reference chosen for vison. 0, 1, 2, 3 four sublattice of vison. The x and y axis of triangular lattice.}
\label{fig:dimer}
\end{figure}

Motivated by the mapping in Eq.~\eqref{eq:Jss=n},
we construct vison creation/annihination operators in the QDM from the string operator in Eq.~\eqref{eq:vstr}.
The $\mathbb Z_2$ spin liquid phase maps to the paramagnetic phase in the FFTFIM,
which is realized in the strong-$\Gamma$ limit, $\Gamma\gg |J_{IJ}|$.
The ground state in this limit has all spins pointing to the $x$ direction, $s_I^x=+1$.
A vison excitation is a flipped spin, $s_I^x=-1$.
Hence, the vison creation/annihilation operator is $s_I^z$.
Note that the creation and annihilation operator is the same, because visons obey a $\mathbb Z_2$ fusion rule and they are their own antiparticle.
The relation in Eq.~\eqref{eq:Jss=n} then implies that the dimer-parity operator $(-1)^{n_{ij}}$ is a product of two vison creation/annihilation operators: it creates/annihilates two visons on the two neighboring dual sites, or moves a vison
between them.
Using this relation repeatedly, we can map the two-point operator $V_{IJ}= s_I^zs_J^z$ to a string of dimer-parity operators.
We first choose an arbitrary string $L:I\rightarrow J$ connecting the two dual sites $I$ and $J$.
Following $L$, we rewrite the correlation function as a string operator,
\[V_{IJ}=\prod_{\langle IJ\rangle\in L} s_I^zs_J^z.\]
Using the relation Eq.~\eqref{eq:Jss=n}, we rewrite this using the dimer-parity operators on the bonds $\langle ij\rangle$ the string $L$ cuts through,
\begin{equation}
	\label{eq:vstr2}
	V_{IJ} = \prod_{\langle ij\rangle\in L}(-1)^{n_{ij}+n_{ij}'},
\end{equation}
where $n_{ij}'=1$ or 0 is determined by $(-1)^{n_{ij}'} = J_{IJ}$.
Here, $n_{ij}'$ can be viewed as a reference dimer configuration, determined by the gauge choice of $J_{IJ}$, as shown in Fig.~S1(c).
Comparing to the vison string operator in Eq.~\eqref{eq:vstr}, the two operators give the same physical result, because they only differ by a static sign independent of the physical dimer configuration.
However, the operator in Eq.~\eqref{eq:vstr2} has the advantage that it has no dependence on the choice of the path $L$, whereas the operator Eq.~\eqref{eq:vstr} has a weak dependence on $L$: the operators defined by two paths $L$ and $L'$ is the same (differ by a minus sign) if the area enclosed by $L$ and $L'$ encloses even (odd) number of sites, respectively.
The independence of the choice of paths makes the operator in Eq.~\eqref{eq:vstr2} convenient for studying the position-dependence and the Fourier transform of $V_{IJ}$.
Next, we introduce a vison creation/annihilation operator in the QDM.
In the FFTFIM, this operator can be obtained by multiplying the two-point operator $s_I^zs_0^z$ with $s_0^z$, where the site $J=0$ is an arbitrary reference point.
Since the mapping from dimer to spin configurations is two-to-one, one can choose $s_0^z=\pm1$ arbitrarily and obtain the following vison creation/annihilation operator in the QDM,
\begin{equation}
	\label{eq:vison}
	v_I = V_{I0}\eta,
\end{equation}
where $V_{I0}$ is a string operator defined in Eq.~\eqref{eq:vstr2} connecting the dual site $I$ to the reference dual site $J=0$, and $\eta=s_J^z=\pm1$ is an arbitrarily chosen sign.
Using this vison operator, it is easy to verify that the vison string operator can be expressed as
\begin{equation}
	\label{eq:vstr3}
	V_{IJ} = v_Iv_J,
\end{equation}
as expected.
The choice of $\eta=\pm1$ does not affect the calculation of any correlation functions, but one need to count both choices to obtain a histogram of the hidden order parameter, as we will discuss later.

\section{Order parameter}
\label{sec:order}

We now construct the order parameter of the vison-condensation phase transition.
Following Ref.~\cite{MoessnerSondhi2001b}, we begin by studying the dispersion relation of the vison excitations in the $\Gamma\gg J$ limit.
In the limit of $J=0$, the vison excitation is a single flipped spin $s_I^x=-1$, with energy gap $\Gamma$.
In the next order,
the Ising coupling $-J_{IJ}s_I^zs_J^z$ can be viewed as a hopping Hamiltonian for the visons,
and gives them a dispersion.
After Fourier transformation, it can be written in the following matrix form,
\begin{equation}
	\label{eq:HJk}
	H_J = \sum_{ka}s_q^{z\dagger} H(q) s_q^z,
\end{equation}
where $s_q^z$, the Fourier transform of $s_I^z$, is a four-component vector: $s_q^z=\begin{pmatrix}
s_{q0}^z & s_{q1}^z & s_{q2}^z & s_{q3}^z
\end{pmatrix}^T$.
Here, the four components $a=0,\ldots,4$ labels the four sublattices, as shown in Fig.~S1(c).
(The dual honeycomb lattice originally has two sublattices, but the pattern of fully frustrated $J_{IJ}$ further doubles the unit cell.)
If we label dual-lattice sites $I$ by the combination of $\bar I$ and $a$, which labels the unit cell and the sublattice, respectively, the Fourier transform can be expressed as
\begin{equation}
	\label{eq:sIk}
	s_{qa}^z = \sum_{\bar I}s_{\bar Ia}^ze^{-i\bm r_{\bar I}\cdot\bm q}.
\end{equation}
Here, the Fourier transform is performed using the positions of the unit cell, $\bm r_{\bar I}$, instead of the position of the actual site.
In this convention, the matrix $H(\bm q)$ has the following form,
\begin{equation}       
	\label{eq:Hq}
	\left(                 
	\begin{array}{cccc}   
		0 & -1+e^{-iq_y} & 0 & -e^{i2q_x}\\  
		-1+e^{iq_y} & 0 & -1 & 0\\  
		0 & -1 & 0 & -1-e^{-iq_y}\\
		-e^{-i2q_x} & 0 & -1-e^{iq_y} & 0\\
	\end{array}
	\right)                 
\end{equation}
where $q_x$ and $q_y$ are components along the basis dual to the real-space unit vectors shown in Fig.~S1(c).

Next, we diagonalize the matrix $H(\bm q)$.
Its eigenvalues reveal the vison dispersion. At each momentum $\bm q$, there are four eigenvalues $E_{q\alpha}$ and four eigenvectors $u_{q\alpha}$.
The quadratic Hamiltonian in Eq.~\eqref{eq:Hq} can be diagonalized using them as the following,
\begin{equation}
	\label{eq:HqD}
	H = \sum_{q\alpha}E_{q\alpha}\tilde s_{-q\alpha}\tilde s_{q\alpha},
\end{equation}
where $\tilde s_{q\alpha}$ is related to $s_{qa}^z$ by the unitary transformation
\begin{equation}
	s_{qa}^z = \sum_{\alpha}u_{q\alpha,a}\tilde s_{q\alpha}.
\end{equation}
Therefore, the operators $\tilde s_{q\alpha}$ represents the vison modes in this model.
In particular, we focus on the lowest band, which we denote by $\alpha=0$.
The dispersion $E_{q0}$ has minima at four momenta: $\bm Q_{0,2}=\pm\left(\frac\pi6,\frac\pi6\right)$ and $\bm Q_{1,3}=\pm\left(-\frac{\pi}6,\frac{5\pi}6\right)$, the $B$ point in the BZ as shown in Fig.1 (b) of the main text, where
the minimum energy is $E_{Q_i}=\Gamma - \frac{\sqrt 6}2 J$.
The eigenvectors at these momenta are
\begin{equation}
	\label{eq:u1234}
	\begin{split}
		u_{Q_00}=u_{Q_20}^\ast &= \begin{pmatrix}
			\frac1F e^{i\frac{\pi}{3}} & \frac1F e^{-i\frac{\pi}{12}}
			& e^{-i\frac\pi{12}} & 1
		\end{pmatrix}^T,\\
		u_{Q_10}=u_{Q_30}^\ast &= \begin{pmatrix}
			e^{-i\frac\pi{3}} & e^{-i\frac{5\pi}12} &
			\frac1F e^{-i\frac{5\pi}{12}} & \frac1F
		\end{pmatrix}^T.
	\end{split}
\end{equation}

At the vison-condensation transition, the vison condenses at the four modes above.
Therefore, the order parameter of this phase transition is the expectation values of the corresponding $\tilde s_{Q_i0}$ operators.
Furthermore, in order to construct a Ginzburg-Landau theory with a position-dependent order-parameter field, we also consider momenta close to but not exactly at $\bm Q_i$.
Hence, we introduce the following fields,
\begin{equation}
	\label{eq:phi-k}
	\phi_i(\bm k) = \tilde s_{Q_i+k, 0}.
\end{equation}
Here, $k\ll1$ is a small momentum.
Fourier-transforming $\phi_i(\bm k)$ yields the real-space order parameter $\phi(\bm r)$.
Here, $\bm r$ still labels the location of the unit cell, as $\bm k$ is a momentum in the Brillouin Zone.
Hence, we denote the real-space order parameter as $\phi_i(\bar I)$, where $\bar I$ labels the unit cell.
\begin{equation}
	\label{eq:phi-x}
	\phi_i(\bar I) = \sum_k\tilde s_{Q_i+k,0}e^{i\bm k\cdot\bm r_{\bar I}}
	= \sum_au_{Q_i0,a}^\ast e^{-i\bm Q_i\cdot\bm r_{\bar I}}s_{\bar Ia}^z.
\end{equation}
Plugging in Eq.~\eqref{eq:u1234}, we get
\begin{equation}
	\label{eq:phi12}
	\begin{split}
		\phi_0(\bar I)=\phi_2(\bar I)^\ast
		&=\left(\frac1F e^{i\frac{\pi}{3}} s_{\bar I0}^z+
		\frac1F e^{-i\frac\pi{12}}s_{\bar I1}^z+
		e^{-i\frac\pi{12}}s_{\bar I2}^z+ s_{\bar I3}^z \right)
		e^{-i\frac\pi6 2\bar x-i\frac\pi6\bar y},\\
		\phi_1(\bar I)=\phi_3(\bar I)^\ast
		&=\left(
		e^{-i\frac\pi{3}}s_{\bar I0}^z+
		e^{-i\frac{5\pi}{12}}s_{\bar I1}^z+
		\frac1F e^{-i\frac{5\pi}{12}}s_{\bar I2}^z+
		\frac1F s_{\bar I3}^z
		\right)e^{i\frac{\pi}6 2\bar x-i\frac{5\pi}6\bar y}
	\end{split}
\end{equation}
Replace the spin operators $s_{\bar Ia}^z$ by the vison operators in Eq.~\eqref{eq:vison}, we can compute the order parameters $\phi_i$ in QDM as
\begin{equation}
	\label{eq:phi12}
	\begin{split}
		\phi_0(\bar I)=\phi_2(\bar I)^\ast
		&=\left(\frac1F e^{i\frac{\pi}{3}} v_{\bar I0}+
		\frac1F e^{-i\frac\pi{12}}v_{\bar I1}+
		e^{-i\frac\pi{12}}v_{\bar I2}+ v_{\bar I3} \right)
		e^{-i\frac\pi6 2\bar x-i\frac\pi6\bar y},\\
		\phi_1(\bar I)=\phi_3(\bar I)^\ast
		&=\left(
		e^{-i\frac\pi{3}}v_{\bar I0}+
		e^{-i\frac{5\pi}{12}}v_{\bar I1}+
		\frac1F e^{-i\frac{5\pi}{12}}v_{\bar I2}+
		\frac1F v_{\bar I3}
		\right)e^{i\frac{\pi}6 2\bar x-i\frac{5\pi}6\bar y}
	\end{split}
\end{equation}
Again, we emphasize that, to verify the emergent O(4) symmetry in $\phi_i$, we need to make a histogram using both $\eta=\pm1$ in Eq.~\eqref{eq:vison}.

\section{Symmetry Transformation of Order parameter}
The O(4) order parameter discussed in the previous section $\phi_i, i=0,1,2,3$, can transform under translation operation $(Tx, Ty)$, mirror operation $(M)$, and rotation operation $(R)$symmetry operation, we list the corresponding $4\times 4$ group representations below.
\begin{equation}       
	Tx=\left(                 
	\begin{array}{cccc}   
		0 & 0 & 0 & e^{-i\frac{5\pi}{12}}\\  
		0 & 0 & e^{-i\frac{\pi}{12}} & 0\\  
		0 & e^{i\frac{5\pi}{12}} & 0 & 0\\
		e^{i\frac{\pi}{12}} & 0 & 0 & 0\\
	\end{array}
	\right)                
	\label{Tx}
\end{equation}
gives new order parameters after the configurations shift one unit along x-axis.

\begin{equation}       
	Ty=\left(                 
	\begin{array}{cccc}   
		e^{-i\frac{\pi}{6}} & 0 & 0 & 0\\  
		0 & e^{-i\frac{5\pi}{6}} & 0 & 0\\  
		0 & 0 & e^{i\frac{\pi}{6}} & 0\\
		0 & 0 & 0 & e^{i\frac{5\pi}{6}}\\
	\end{array}
	\right)                
	\label{Ty}
\end{equation}
is the transformation of order parameter after y-axis translation.

\begin{equation}       
	M=\frac{1}{\sqrt{2}}\left(                 
	\begin{array}{cccc}   
		0 & e^{i\frac{5\pi}{6}} & e^{i\frac{\pi}{12}} & 0\\  
		e^{-i\frac{5\pi}{6}} & 0 & 0 & e^{-i\frac{7\pi}{12}}\\  
		e^{-i\frac{\pi}{12}} & 0 & 0 & e^{-i\frac{5\pi}{6}}\\
		0 & e^{i\frac{7\pi}{12}} & e^{i\frac{5\pi}{6}} & 0\\
	\end{array}
	\right)                
	\label{M}
\end{equation}
M is a mirror operator, the mirror axis is perpendicular to the x-axis.

\begin{equation}       
	R=\frac{1}{\sqrt{2}}\left(                 
	\begin{array}{cccc}   
		0 & e^{i\frac{\pi}{12}} & e^{i\frac{5\pi}{6}} & 0\\  
		e^{-i\frac{7\pi}{12}} & 0 & 0 & e^{-i\frac{5\pi}{6}}\\  
		e^{-i\frac{5\pi}{6}} & 0 & 0 & e^{-i\frac{\pi}{12}}\\
		0 & e^{i\frac{5\pi}{6}} & e^{i\frac{7\pi}{12}} & 0\\
	\end{array}
	\right)                
	\label{R}
\end{equation}
is the $C_6$ rotation.

It can also be seen from above analysis that the order parameter is distributed on a four-dimensional sphere. As shown in the Fig.~3 in the main text, at the QSL-VBS transition point, the O(4) order parameter lives homogeneous on the sphere hence acquires with emergent O(4) symmetry; and deep inside the $\sqrt{12}\times\sqrt{12}$ VBS phase, the order parameters concentrate on few discrete points on the sphere, hence suggesting the O(4) symmetry breaking.
We notice that the positions of the peaks in the order-parameter distribution in Fig.~3(c,d) disagree with the theoretical prediction based on the Landau theory in Ref.~\cite{MoessnerSondhi2001b}:
in particular, the Landau theory puts peaks at angles $\theta_{a,b}$ that are multiples of $\pi/24$, where the peaks we observe in Fig. 3(c,d) are localed at angles of $(2n-1)\pi/48$.
We think this disagreement may be because the Landau theory, with the lowest-order anisotropies, does not apply to parameters like $V=0$, which are far away from the quantum critical point.
We leave a more detailed study of peak locations near the quantum critical point to future works.

\section{Fitting excitation gaps}
In order to compare the size of the excitations gaps as shown in the Fig.~2 of the main text, one can directly fit the exponential decay as $\exp(-\Delta\times\tau)$ from the imaginary time correlation functions of $C_{d}(\mathbf{q},\tau)$, $C_{v}(\mathbf{q},\tau)$ and $C^{vc}_{d}(\mathbf{q},\tau)$. Here we take the data of $V=1$, $L=12$, and $\beta=200$ as an example to fit the imaginary time correlation functions at different momenta, the results are shown in Fig.~S2.

\begin{figure}[htp]
	\includegraphics[width=0.7\columnwidth]{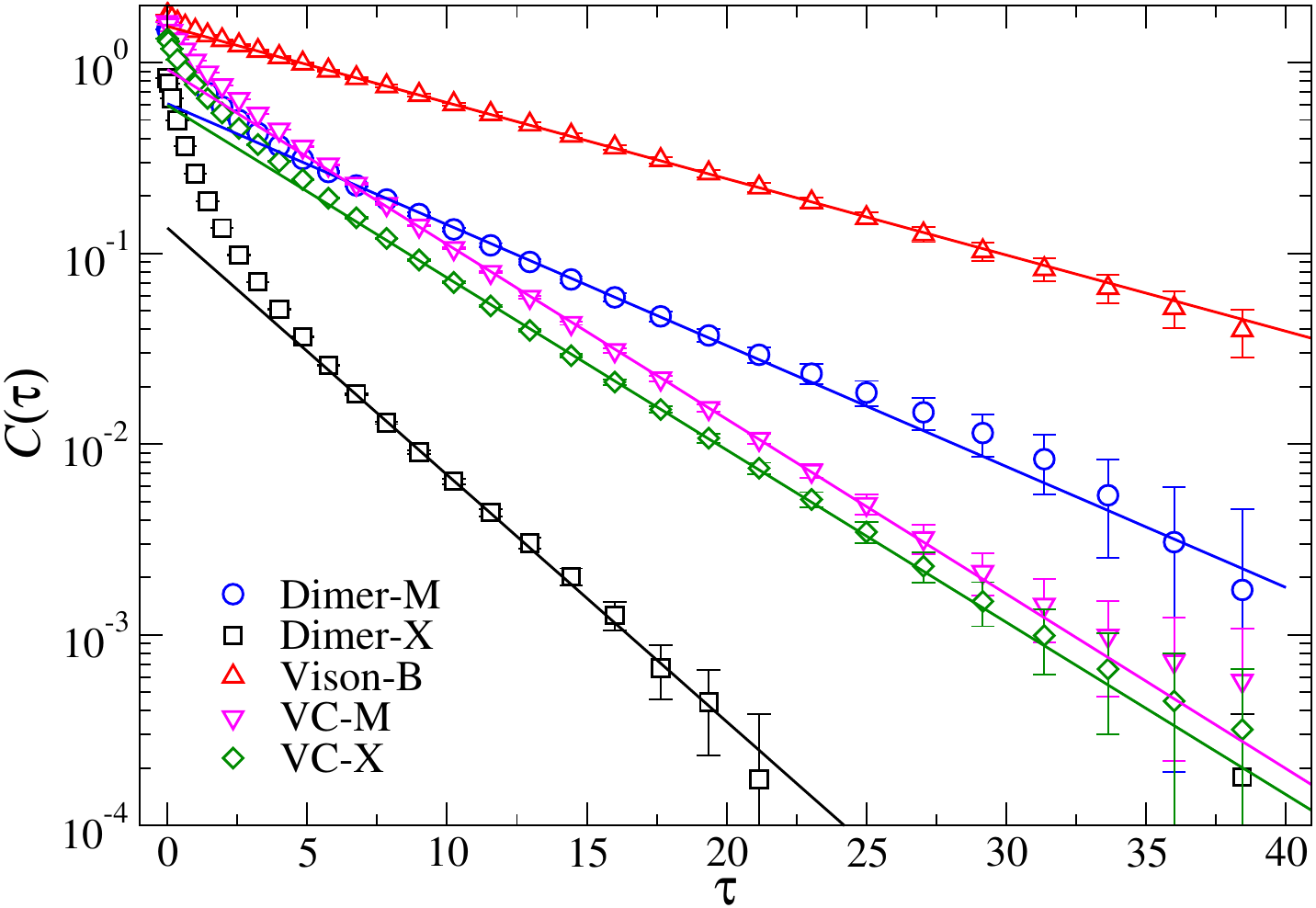}
\caption{With the form of $Ae^{-\Delta\times\tau}$, we fit the excitation gaps $\Delta$ for the data with parameters $V=1$, $L=12$, and $\beta=200$. For $C_{d}(\mathbf{q},\tau)$ at $M$ and $X$ point, the obtained gaps $\Delta$ are about 0.15(1) and 0.30(1), the amplitudes A are $0.61(2)$ and $0.14(1)$. For $C_{v}(\mathbf{q},\tau)$ at $B$ point, $\Delta=0.09(2)$ and $A=1.55(1)$. And for vison-convolution $C^{vc}_{d}(\mathbf{q},\tau)$ at $M$ and $X$, the obtained gaps $\Delta$ are 0.21(1) and 0.21(1), amplitudes A are $0.92(2)$ and $0.60(2)$.}
	\label{fig:gap}
\end{figure}

It is clear from such analysis, that the vison convolution gap at $M$ point is about twice of the single vison gap at $B$ point, and the dimer gap, which represents the bounding of the two visons due to interaction, is smaller than twice of the single vison gap. Similar analysis with more system sizes and finite size scaling, due to the huge computation burden for dynamic measurements, are under preparation for future works.

\end{document}